\documentclass[]{aastex631}

\shorttitle{Persistent Upflows and Downflows at Active Region boundaries Observed by SUTRI and AIA}
\shortauthors{Yuchuan Wu et al.}

\graphicspath{{./}{figures/}}

\def \kms {{\rm km\;s$^{-1}$}}
\def \nevii {Ne\,{\sc vii}}

\begin{document}

\title{Persistent Upflows and Downflows at Active Region boundaries Observed by SUTRI and AIA}

\author{Yuchuan Wu}
\affiliation{National Astronomical Observatories, Chinese Academy of Sciences, Beijing, 100101, China}
\affiliation{School of Astronomy and Space Science, University of Chinese Academy of Sciences, Beijing 100049,  China}

\author{Zhenyong Hou}
\affiliation{School of Earth and Space Sciences, Peking University, Beijing, 100871, China}

\author{Wenxian Li}
\affiliation{National Astronomical Observatories, Chinese Academy of Sciences, Beijing, 100101, China}
\affiliation{Key Laboratory of Solar Activity and Space Weather, National Space Science Center, Chinese Academy of Sciences, Beijing 100190, China}

\author{Xianyong Bai}
\affiliation{National Astronomical Observatories, Chinese Academy of Sciences, Beijing, 100101, China}
\affiliation{Key Laboratory of Solar Activity and Space Weather, National Space Science Center, Chinese Academy of Sciences, Beijing 100190, China}
\affiliation{School of Astronomy and Space Science, University of Chinese Academy of Sciences, Beijing 100049,  China}

\author{Yongliang Song}
\affiliation{National Astronomical Observatories, Chinese Academy of Sciences, Beijing, 100101, China}
\affiliation{Key Laboratory of Solar Activity and Space Weather, National Space Science Center, Chinese Academy of Sciences, Beijing 100190, China}

\author{Xiao Yang}
\affiliation{National Astronomical Observatories, Chinese Academy of Sciences, Beijing, 100101, China}
\affiliation{Key Laboratory of Solar Activity and Space Weather, National Space Science Center, Chinese Academy of Sciences, Beijing 100190, China}

\author{Ziyao Hu}
\affiliation{National Astronomical Observatories, Chinese Academy of Sciences, Beijing, 100101, China}
\affiliation{School of Astronomy and Space Science, University of Chinese Academy of Sciences, Beijing 100049,  China}

\author{Yuanyong Deng}
\affiliation{National Astronomical Observatories, Chinese Academy of Sciences, Beijing, 100101, China}
\affiliation{School of Astronomy and Space Science, University of Chinese Academy of Sciences, Beijing 100049,  China}
\affiliation{Key Laboratory of Solar Activity and Space Weather, National Space Science Center, Chinese Academy of Sciences, Beijing 100190, China}

\author{Kaifan Ji}
\affiliation{Yunnan Observatories, Chinese Academy of Sciences, Kunming, 650011, China}

\correspondingauthor{Zhenyong Hou; Xianyong Bai}
\email{zhenyonghou@pku.edu.cn; xybai@bao.ac.cn}

\begin{abstract}
{Upflows and downflows at active region (AR) boundaries have been frequently observed with spectroscopic observations at extreme ultraviolet (EUV) passbands.
In this paper, we report the coexistence of upflows and downflows at the AR boundaries with imaging observations from the Solar Upper Transition Region Imager (SUTRI) and the Atmospheric Imaging Assembly (AIA). 
With their observations from 2022 September 21 to 2022 September 30, we find 17 persistent opposite flows occurring along the AR coronal loops.
The upflows are prominent in the AIA 193\,\AA\ images with a velocity of 50-200 \kms, while the downflows are best seen in the SUTRI 465\,\AA\ and AIA 131\,\AA\ images with a slower velocity of tens of kilometers per second (characteristic temperatures (log $T$(K)) for 193\ \AA, 465\ \AA \ and 131\ \AA \ are 6.2, 5.7, 5.6, respectively).
We also analyze the center-to-limb variation of the velocities for both upflows and downflows.
The simultaneous observations of downflows and upflows can be explained by the chromosphere-corona mass-cycling process, in which the localized chromospheric plasma is impulsively heated to coronal temperature forming a upflow and then these upflows experience radiative cooling producing a downflow with the previously heated plasma returning to the lower atmosphere.
In particular, the persistent downflows seen by SUTRI provide strong evidence of the cooling process in the mass cycle.}
\textcolor{black}{For upflows associated with open loops, part of the plasma is able to escape outward and into the heliosphere as solar wind.}
\end{abstract}


\section{Introduction} \label{sec:intro}

From EUV and X-ray imaging observations, quasi-periodic propagating disturbances (PDs) along the fan-structures at active region (AR) boundaries have been observed for decades \textcolor{black}{\citep{1999SoPh..186..207B}}. 
The upward propagating disturbances (UPDs) are usually quite common and have been observed in polar plumes \citep[e.g.,][]{1997ApJ...491L.111O,1999ApJ...514..441O,1998ApJ...501L.217D,2011ApJ...736..130T,2015ApJ...809L..17J,2016RAA....16...93J} and ARs \textcolor{black}{\citep[e.g.,][]{1999SoPh..186..207B,2000A&A...355L..23D,2011A&A...526A..58S,2013ApJ...778...26U,2021ApJ...918...33H}}. The UPDs always has a fast propagation speed of 50-200 km s$^{-1}$, while downward propagating disturbances (DPDs) are slower with a speed of tens of km s$^{-1}$ \citep{2012ApJ...749...60M}. A temperature dependence of the existence of UPDs or DPDs and their velocities is observed at AR boundaries. 
For example, \citet{2011A&A...532A..96K} and \citet{2012ApJ...749...60M} found that the UPDs appear in the hot 193 \AA~passband with a fast velocity of about 100 km s$^{-1}$, while the sporadic slow DPDs are formed in the cool passbands such as 131 \AA~with a speed of 15 km s$^{-1}$. 
\citet{2012ApJ...749...60M} also found the coexistence of upflows and downflows in the 171 \AA~observations.

Blue and red shifts of spectral lines, which usually coexist with PDs, have been frequently reported from spectroscopic observations \citep[e.g.,][]{2010ApJ...722.1013D,2011ApJ...738...18T,2012ApJ...748..106T}.
The corresponding velocities can be estimated from the blue and red shifts.
\citet{2011ApJ...738...18T} applied the single Gaussian fit to the observed coronal line profiles and yielded a blue shift of 10-50 km s$^{-1}$.
However, a more detailed analysis using double-Gaussian fitting technique \citep[e.g.,][]{2010A&A...521A..51P,2015ApJ...805...97K} and red-blue (RB) asymmetry analysis \citep[e.g.,][]{2009ApJ...701L...1D,2010ApJ...722.1013D} suggests that the
coronal emission consist of two components, i.e. a primary component for stationary background and a secondary component associated with high-speed upflows \citep{2011ApJ...738...18T,2012ApJ...760L...5B}.
The primary component is often blueshifted but with a slower speed of about 10 km s$^{-1}$, 
while the secondary component corresponds to a much faster velocity of 50--150 km s$^{-1}$,  sometime reaching 200 km s$^{-1}$ \citep{2010ApJ...722.1013D,2011ApJ...738...18T}. 
In addition, a clear center-to-limb variation of doppler shifts is observed \citep{2012ApJ...748..106T,2019ApJ...886...46G,2023ApJ...944..158R}. 
The doppler shifts are close to 0 when the AR is located at the limb of the Sun and reach maximum as the AR rotates to the center of solar disk \citep{2019ApJ...886...46G,2023ApJ...944..158R}.
The dependence of the red and blue shifts on formation temperature has also been studied.
An average red shift is normally observed in spectral lines with lower formation temperatures and blue shift for lines with higher formation temperatures \citep[e.g.,][]{1999ApJ...522.1148P,2009ApJ...694.1256T,2011A&A...534A..90D,2004A&A...424.1025X,2009SoPh..255..119R,2014ApJ...794..109F}. 
\citet{2004A&A...424.1025X} found that the transition from red shift to blue shifts occurs at a temperature (log $T$(K)) of 5.8 for coronal holes and 6.0 for quiet Sun. For AR fan-like structures, the transition temperature (log $T$(K)) is 6.0 \citep{2011ApJ...727...58W}. 

DPDs and red shifts are explained by downflows \citep{2012ApJ...749...60M}, while there are two explanations for UPDs and blue shifts, upflows \citep[e.g.,][]{2007Sci...318.1585S,2008ApJ...676L.147H,2011ApJ...738...18T} and slow-mode magneto-acoustic waves \citep[e.g.,][]{2009ApJ...696.1448W,2009SSRv..149...65D,2011ApJ...737L..43N}, and now researchers believe that both upflows and waves exist \citep{2021SoPh..296...47T}.
Thus co-spatial UPDs and DPDs (labeled as bidirectionally propagating disturbances, BPDs, hereafter) can be explained by co-spatial upflows and downflows, and their temperature-dependence imply a chromosphere-corona mass cycle process.
As suggested by \citet{2012ApJ...749...60M}, the plasma in the chromosphere is impulsively heated to transition region and coronal temperature, exhibiting as high-speed upflows. 
Then these upflows experience radiative cooling process and the previously heated materials will slowly return to the chromosphere, producing slow downflows. 
A similar scenario of mass cycling has also been proposed by \citet{2008ApJ...685.1262M} and \citet{2012ApJ...744...14Y}.
\textcolor{black}{Additionally, for UPDs in open loops associated with open-field lines, part of the plasma can escape outward and into the heliosphere as fast solar wind. For example, \citet{2023ApJ...955L..38U} reported self-similar outflows of hundreds of km s$^{-1}$ above a polar coronal hole, and \citet{2023ApJ...943..156K} reported several jets in AR during a failed eruption. These events are believed to be a mixture of hot plasma and Alfv$\rm \acute{e}$n waves, and can be a source of fast solar wind \citep{2023ApJ...945...28R,2023ApJ...955L..38U,2022ApJ...933...21K,2023ApJ...943..156K}.}
Thus the study of the BPDs are highly related to our understanding of the mass and energy transport mechanism in the solar atmosphere as well as the formation of the upper atmospheric heating.

By the analyses of observations from Atmospheric Imaging Assembly \citep[AIA,][]{2012SoPh..275...17L} and the Solar Upper Transition Region Imager \citep[SUTRI,][]{2023RAA....23f5014B,2023RAA....23i5009W}, we present the spatial and temporal properties of the upflows and downflows occurring at the AR boundaries and their appearance in the hot and cool passpands. 
In particular, SUTRI provides full-disk solar observations at Ne VII 465 \AA~line formed at a temperature of 0.5 MK in transition region, which has been rarely explored. 
In Section\,\ref{sec:data} we describe the observations and analysis methods. In Section\,\ref{sec:result} we present the characteristics of the BPD events and discussions.
Finally, we summarize our findings in Section\,\ref{sec:sum}.

\section{Observations and Analysis}
\label{sec:data}

In this research, we mainly analyzed the imaging observations taken by SUTRI on board the Space Advanced Technology demonstration satellite (SATech-01) and the AIA on board the Solar Dynamics Observatory \citep[SDO,][]{2012SoPh..275....3P}.
SUTRI provides narrow-band imaging of full solar disk in the \nevii\,465\,\AA\ passband, which is formed at a rarely sampled temperature of about 0.5 MK \cite[log $T$(K)$\approx$5.7,][]{2017RAA....17..110T}.
The normal cadence of SUTRI observations is 30 s and the sampling resolution is $1.22^{\prime\prime}$/pixel.
Since about 1/3 of SATech-01's orbit period ($\sim$96 min) is in the earth eclipse, SUTRI’s maximum continuous observation time is about an hour.
AIA provides full-disk solar observations in seven different EUV passbands, i.e., 94\,\AA~(Fe XVIII), 131\,\AA~(Fe VIII for AR, Fe  XXI for flare), 171\,\AA (Fe IX), 193\,\AA~(Fe XII for AR, Fe XXIV for flare), 211\,\AA~(Fe XIV), 304\,\AA~(He II), and 335\,\AA~(Fe XVI), spanning a temperature range from about $6\times 10^4$ K to $2\times 10^7$ K.
For our analysis, we mainly used 304\,\AA, 131\,\AA, 171\,\AA, and 193\,\AA\ passbands, with characteristic temperatures (log $T$(K)) of 4.7, 5.6, 5.8, and 6.2, respectively.
The AIA images have a sampling resolution of $0.6^{\prime\prime}$/pixel and a cadence of 12 s.

PDs in ARs are sometimes covered by the surrounding dynamic structures and are hard to see in the EUV observations.
To better reveal the faint propagating signatures, \citet{2011ApJ...736..130T} and \citet{2012ApJ...749...60M} used the detrended intensities for their analyses.
Following the method of \citet{2011ApJ...736..130T} and \citet{2012ApJ...749...60M}, we obtained the detrended intensities by subtracting a $X$-minute running average from the raw intensities. 
To highlight PDs with different velocities \citep{2012ApJ...749...60M}, we choose a $X$ of 8 and 20, respectively, to enhance the relative contribution of the faster and slower PDs to the total emissions. 

We analyze SUTRI and AIA observations of AR 13105, 13106, and 13107 from 2022 September 21 to 2022 September 30, during which time the ARs rotate from eastern limb to western limb. By studying the detrended movies, we can easily discern the persistent UPDs or/and DPDs occurring along the AR coronal loops depending on the temperature of the diagnostic passband used.
For event with simultaneous appearance of UPDs and DPDs on the same structure, we define it as BPD and include it in our studies.
Then we perform the space–time analysis to investigate the evolution of the PDs along the coronal loop for the four AIA passbands (304, 131, 171, and 193 \AA) and SUTRI observations (465 \AA).

\begin{deluxetable*}{cccccccccc}
\label{table1}
\tablecaption{Detailed information of the BPDs.}
\tablewidth{0pt}
\tablehead{
BPD & Time & Location & \multicolumn{2}{c}{Velocity (\kms)} & SUTRI & \multicolumn{4}{c}{AIA} \\
ID & (UT) & (arcsec) & DPD & UPD & 465\,\AA & 304\,\AA & 131\,\AA & 171\,\AA & 193\,\AA 
}
\startdata
1 & 2022-9-21 00:19--01:10 & (-796,-264) & \textcolor{black}{35$\pm$7} & \textcolor{black}{171$\pm$30} & d & & d & u & u  \\
2 & 2022-9-21 00:19--01:10 & (-785,-260) & \textcolor{black}{44$\pm$6} & \textcolor{black}{143$\pm$18} & d & d & d & \textcolor{black}{d} & u  \\
3 & 2022-9-21 17:40--18:32 & (-688,-275) & \textcolor{black}{65$\pm$5} & \textcolor{black}{163$\pm$24} & d & d & \textcolor{black}{d,u} & d,u & d,u  \\
4 & 2022-9-23 10:42--11:10 & (-427,-304) & \textcolor{black}{36$\pm$14} & \textcolor{black}{153$\pm$16} & d & & d &  & u \\
5 & 2022-9-24 10:23--11:10 & (-572,-492) & \textcolor{black}{34$\pm$6} & \textcolor{black}{180$\pm$40} & d & & d & d,u & d,u \\
6 & 2022-9-24 19:54--20:37 & (-116,-405) & \textcolor{black}{25$\pm$4} & \textcolor{black}{139$\pm$23} & d & & d & d,u & u  \\
7 & 2022-9-24 21:26--22:11 & (-88,-393) & \textcolor{black}{41$\pm$5} & \textcolor{black}{71$\pm$9} & d & & d & d & u \\
8 & 2022-9-24 23:01--23:46 & (-129,-317) & \textcolor{black}{48$\pm$6} & \textcolor{black}{141$\pm$25} & d & d & d & d & \textcolor{black}{d,u} \\
9 & 2022-9-25 08:26--09:13 & (25,-259) & \textcolor{black}{31$\pm$3} & \textcolor{black}{128$\pm$21} & d & & d &  & d,u \\
10 & 2022-9-25 08:26--09:13 & (24,-259) & \textcolor{black}{29$\pm$3} & \textcolor{black}{113$\pm$16} & d,u & & d,u & d,u & u \\
11 & 2022-9-27 06:13--06:54 & (-57,-452) & \textcolor{black}{38$\pm$6} & \textcolor{black}{113$\pm$18} & d & & d & d & u \\
12 & 2022-9-27 07:47--08:28 & (393,-394) & \textcolor{black}{48$\pm$8} & \textcolor{black}{86$\pm$8} & d & \textcolor{black}{d} & d & d & \textcolor{black}{d,u} \\
13 & 2022-9-27 12:32--12:54 & (475,-349) & \textcolor{black}{31$\pm$9} & \textcolor{black}{100$\pm$15} & d & & d & & u \\
14 & 2022-9-27 17:16--17:55 & (527,-238) & \textcolor{black}{28$\pm$5} & \textcolor{black}{236$\pm$40} & d & d & d & & u \\
15 & 2022-9-27 23:34--9-28 00:14 & (527,-238) & \textcolor{black}{48$\pm$7} & \textcolor{black}{104$\pm$10} & d & & d & d & u \\
16 & 2022-9-28 05:53--06:48 & (612,-270) & \textcolor{black}{32$\pm$6} & \textcolor{black}{149$\pm$18} & d & d & d & & u \\
17 & 2022-9-29 05:34--06:28 & (725,-248) & \textcolor{black}{19$\pm$3} & \textcolor{black}{64$\pm$8} & d & d & d & & u  
\enddata
\tablecomments{
We use the coordinates of the footpoints of the AR loops as the locations of the BPDs and give them in the third column.
In the last five columns, `d' and `u' represent, respectively, the presence of a DPD and UPD in each passband.}
\end{deluxetable*}

\begin{figure}[ht!]
\includegraphics[width=1.0\textwidth]{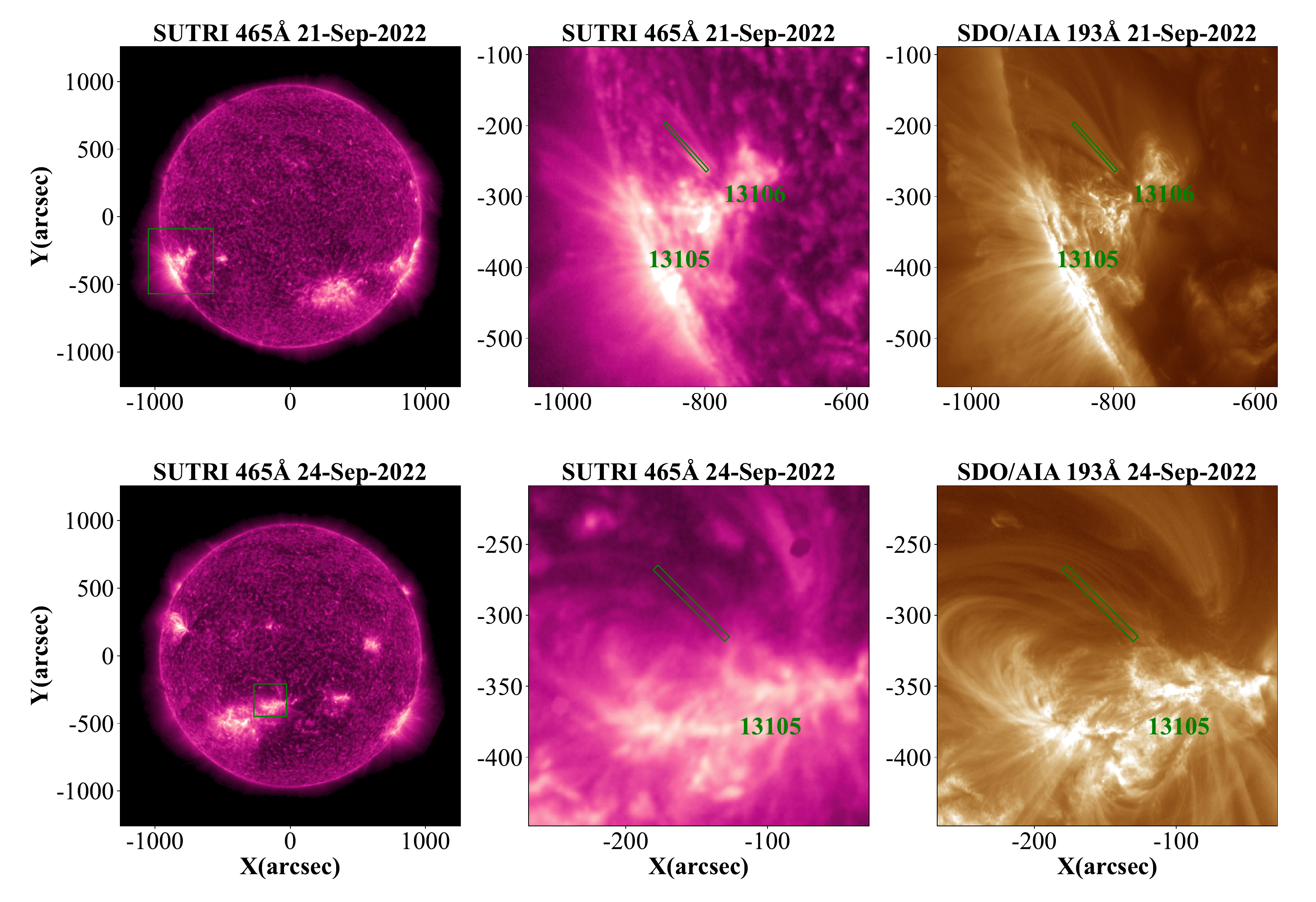}
\caption{Solar observations on 2022 September 21 (top row) and September 24 (bottom row).
The left column shows the full-disk SUTRI images with a square region showing the ARs where the BPDs occur.
The middle and right columns show the zoomed-in images of the square regions in SUTRI 465\,\AA\ and AIA 193\,\AA~passbands, respectively.
The green rectangles outlined in the middle and right panels represent the location of the space–time plots studied in Figures \ref{fig:2} and \ref{fig:3}.\\
(Animations of BPD 1 and BPD 8 are available in the online journal\textcolor{black}{, showing the evolution of these two events. Note that animation of BPD 1 shows observation data of 2022 September 21 00:19-00:54 UT, and animation of BPD 8 shows observation data from 2022 September 24 23:01-23:46 UT. Animation of BPD 1 has a shorter duration compared with BPD 1 in Table 1, because we cut the last frames in which part of SUTRI observations are blocked due to the earth eclipse for a better vision effect. Animation for each event includes four panels, the first row shows original images of SUTRI 465\ \AA\ and AIA 193\ \AA\ passbands, and the second row shows corresponding 20-minute detrended images. The green rectangles outlined in all panels represent the location of the space–time plots studied in Figures \ref{fig:2} and \ref{fig:3}.})
}
\label{fig:1}
\end{figure}

\begin{figure}[ht!]
\includegraphics[width=1.0\textwidth]{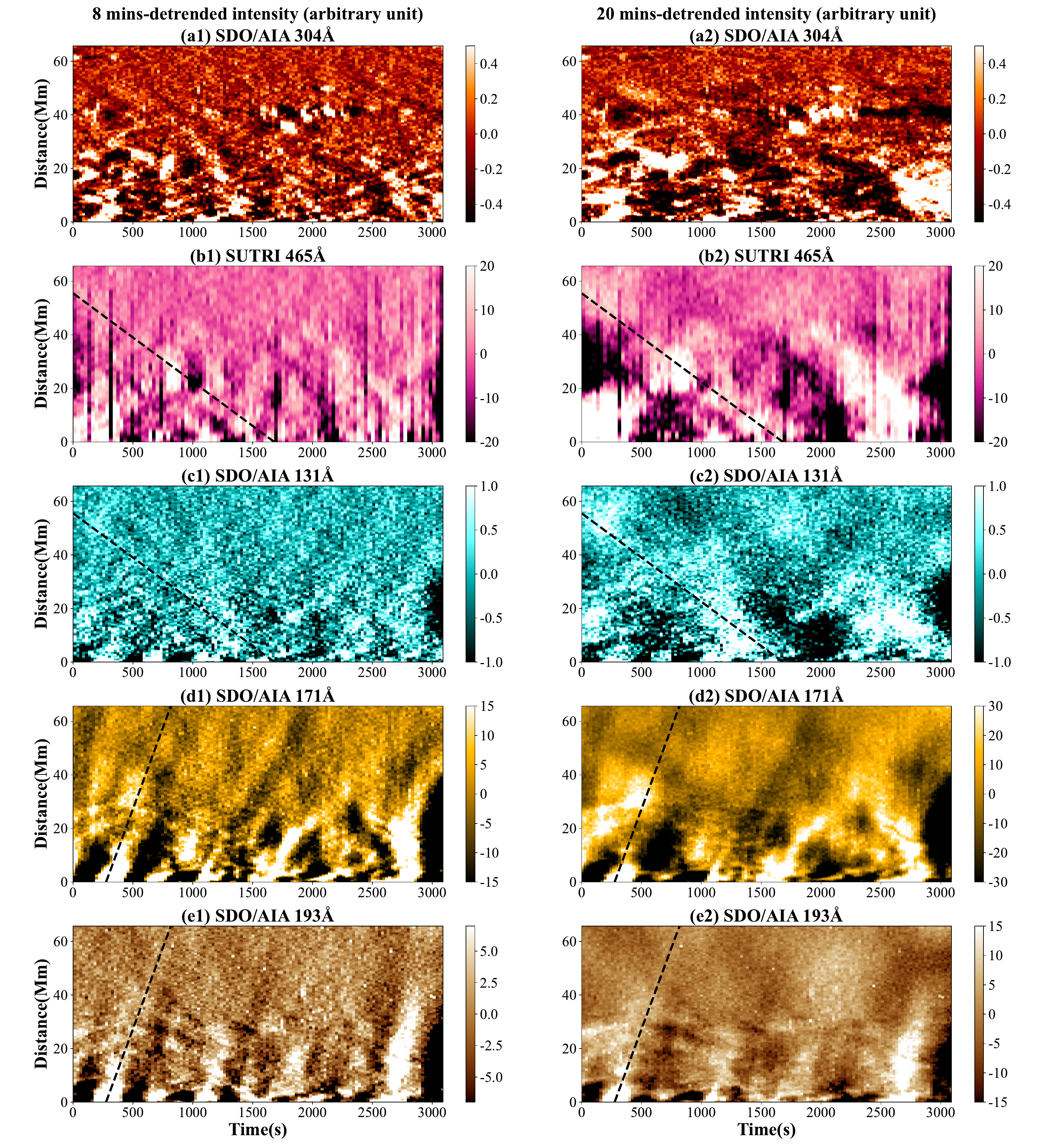}
\caption{From top to bottom: Space–time plots of the detrended intensity signal for BPD\,1 in the AIA 304\,\AA, SUTRI 465\,\AA, AIA 131\,\AA, 171\,\AA, and 193\,\AA\ passbands along the path identified in Figure \ref{fig:1}.
The left and right panels show the 8-minute and 20-minute detrended intensity signals, respectively. The black dashed lines show the identified UPDs and DPDs.}
\label{fig:2}
\end{figure}

\begin{figure}[ht!]
\includegraphics[width=1.0\textwidth]{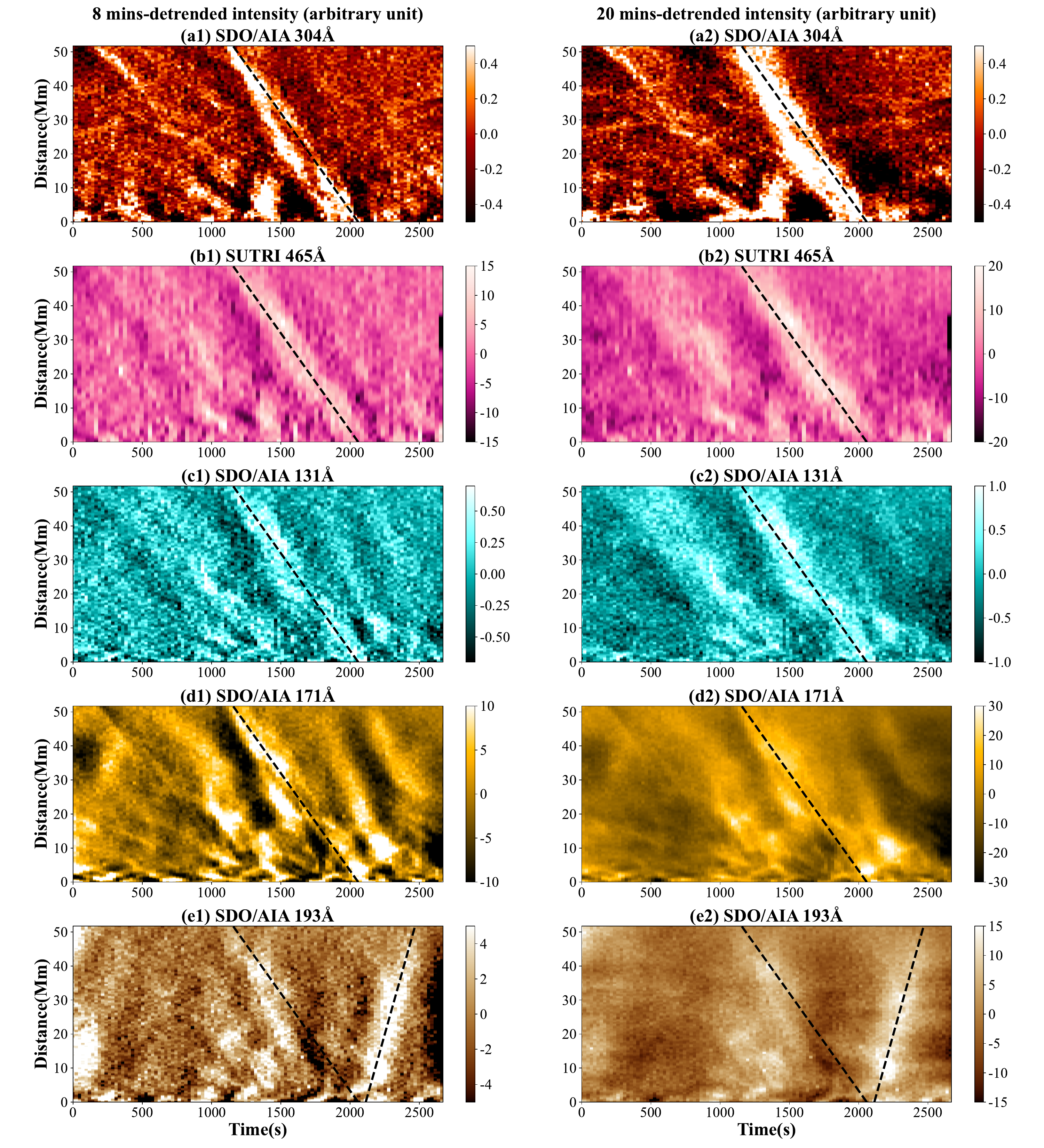}
\caption{Same as Figure\,\ref{fig:2}, but for BPD\,8.}
\label{fig:3}
\end{figure}

\begin{figure}[ht!]
\includegraphics[width=1.0\textwidth]{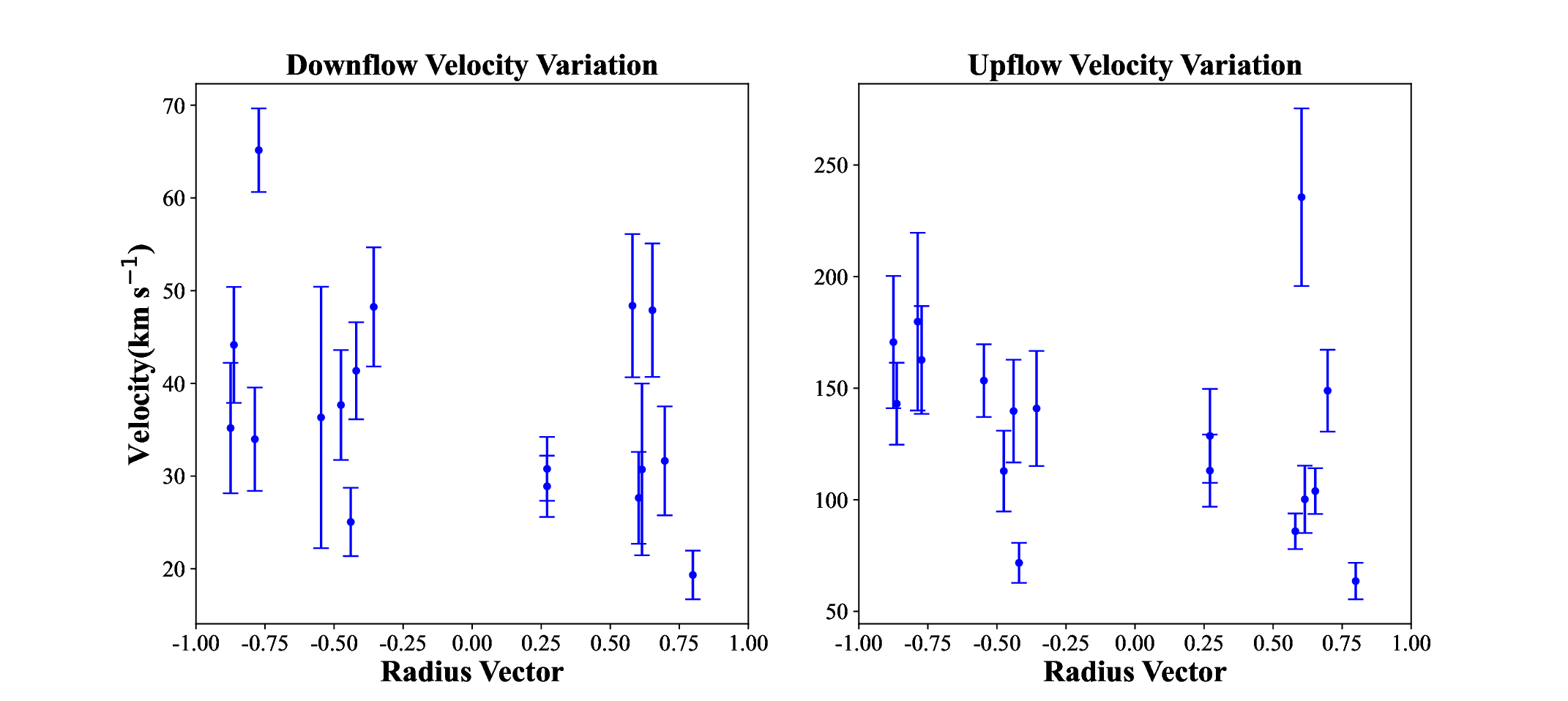}
\caption{Scatterplots of downflow (left panel) and upflow (right panel) velocities as a function of radius vector. \textcolor{black}{The error bars represent the uncertainties of the velocities.}}
\label{fig:4}
\end{figure}

\section{Results and Discussion}
\label{sec:result}
We identify 17 BPD events on the coronal structures and the detailed information, i.e., occuring time, locations, propagating velocities, and appearance in SUTRI and AIA images, are listed in Table \ref{table1}.
The ARs appeared on the east limb on 2022 September 21 and moved to the disk center following the rotation of the Sun on 2022 September 24 and disappeared behind the western limb on September 30.
\textcolor{black}{By observing the loops structure in AIA 193 \AA, we can see that most of the magnetic geometry of the studied loops are open, and there are also several events with closed loops.} Figure\,\ref{fig:1} shows the snapshot of SUTRI and SDO/AIA 193 \AA~observations for BPD 1 (upper panels) and 8 (lower panels).
\textcolor{black}{BPD 1 is an event within an open loop, while BPD 8 is an event within a closed loop.}
From the online supporting movies, we can see clearly the PDs in the fan-like structure at the boundary of the ARs, i.e., slow down-flowing plasma streams in SUTRI 465 Å and fast upward motions in the AIA 193 \AA\ passband.
Note that SUTRI’s maximum continuous observation time is about 60 minutes due to the earth eclipse.

Figure\,\ref{fig:2} and Figure\,\ref{fig:3} show the space-time plots for BPD\,1 and BPD\,8, respectively. 
The starting point of $Y$-axis in the space-time plots corresponds to the footpoint of the AR loops.
The left panels show the signal following the removal of an 8-minute running average from the raw intensity while the right panels show the 20-minute detrended signals.
From the figure we can see the evolution of the PDs along the coronal loop for the AIA 304\,\AA, SUTRI 465\,\AA, 131\,\AA, 171\,\AA, and 193\,\AA\ passbands from top to bottom.
Note that a running average of two frames was applied to AIA images to improve the signal-to-noise ratio, especially for 304 and 131 \AA.
In Figures \ref{fig:2} and \ref{fig:3}, the DPDs and UPDs appear as inclined bright stripes with opposite directions. The left and right panels show clear difference in the patterning visible.
For example, the inclined black dashed lines indicate an apparent downward motion of material along the structure in Figure\,\ref{fig:2} (b2) and (c2), while no obvious signatures are shown in Figure\,\ref{fig:2} (b1) and (c1).
For upflows in AIA 193\,\AA, however, the space-time plots of the 8-minute detrended intensity reveal more signatures than that of the 20-minute detrended intensity.
\textcolor{black}{\citet{2012ApJ...749...60M} have observed the same phenomenon and discussed the explanation. As what is shown in Figure \ref{fig:2} and \ref{fig:3}, upflows are associated with shorter lifetime and downflows with longer lifetime. For a smaller detrending time which is shorter than upflows’ lifetime, a single upflow can be well sampled from beginning to its end. However, only a small portion of a single downflow is sampled during the lifetime of a single upflow. That is because the intensity variation of downflows is removed as the background in the detrended intensity plots. Consequently, for some downflows, they are not so clear or even cannot be observed in the 8-minute detrended plots. While for a longer detrending time which is comparable to upflows lifetime, only a part of the upflows can be sampled from their beginning to the end. So in the 20-minute detrended plots, fewer upflows can be observed. However, for downflows, they can be sampled from beginning to the end, so they can be clearly viewed in 20-minute detrended plots.}

The appearance of the BPDs in each passpand is given in the last five columns of Table\,\ref{table1}, from which 
we found that the downflows are prominent in SUTRI 465\,\AA\ and AIA 131\,\AA\ while the upflows are best seen in AIA 193\,\AA.
In some events, we can see the appearance of both downflows and upflows in the same passband.
The above analysis indicates that the flows are temperature-dependent, with upflow of PDs visible in the hotter passbands while downflows in cooler passbands; this is consistent with previous results by \citet{2012ApJ...749...60M}.

Based on the appearances of the downflows in the EUV passbands, we divided the BPDs into two categories.
In category\,1, downflows only appear in the low-temperature passbands such as the SUTRI 465\,\AA\ and AIA 131\,\AA\ (e.g., BPD\,1).
On the other hand, the BPDs in category\,2 show downflow signatures both in the low- and high-temperature passbands (e.g., BPD\,8 appearing in AIA 193\,\AA).
\textcolor{black}{Category\,1 includes BPD\,1, BPD\,4, BPD\,13, BPD\,14, BPD\,16, and BPD\,17, and category\,2 includes BPD\,2, BPD\,3, BPD\,5, BPD\,6, BPD\,7, BPD\,8, BPD\,9, BPD\,10, BPD\,11, BPD\,12, and BPD\,15.}

The projected velocities of the BPDs on the plane-of-sky (POS) can be estimated by calculating the slopes of the strips, which are indicated by the dashed line in the space-time maps.
\textcolor{black}{When the DPDs/UPDs appeared in multiple passbands, we choose the passband in which DPDs/UPDs can be most clearly observed to estimate POS; they are usually SUTRI 465 \AA\ for DPDs and AIA 193 \AA\ for UPDs. We first define a threshold and select the signals with higher (or lower, in case of a dark stripe) detrended intensity than the threshold in a region where a bright stripe can be observed. 
We choose a suitable threshold to highlight the stripe and separate it from the background.
Then, we use a linear fitting to get the slope as the estimated velocity. The error of velocity mainly comes from the estimate of distance, and is estimated by $\frac{E(\Delta D)}{\Delta T}$, where $\Delta T$ is estimated by the lasting time of PD, and $E(\Delta D)$ is estimate by $\sqrt{2}RMSE(D)$, where $RMSE(D)$ means the root mean square error of the distance in the linear fitting.}
Then if there are multiple DPDs/UPDs in one event, we calculate their average velocity.
\textcolor{black}{As an example, the POS velocities of DPDs in BPD\,1 and BPD\,8, as shown in Figure\,\ref{fig:2} and Figure\,\ref{fig:3}, are usually within the range of 33--57\,\kms, while the UPDs are always associated with a fast speed, i.e. 121--219\,\kms. The averaged POS velocities of DPDs and UPDs in BPD\,1 (BPD\,8) are 35$\pm$7 (40$\pm$6) \,\kms and 171$\pm$30 (141$\pm$25) \,\kms, respectively.}
\textcolor{black}{The estimated POS velocities together with their estimated errors for all the 17 events are shown in Table\,\ref{table1}.
We can see that the downflow usually has a POS velocity of 19--65\,\kms, while the upflow velocity ranges from 64\,\kms\ to 236\,\kms.}

Previous studies have demonstrated that the prevalent UPDs appearing in the fan-like structures at AR boundaries are related to the high-speed secondary components (blueward asymmetries) of coronal lines \citep{2021SoPh..296...47T}.
In this work, the velocity of the upward propagating features is estimated to be \textcolor{black}{64--236}\,\kms, which is consistent with the previous results.
The velocities of the downward propagating features are similar to those obtained from the redshifts of the transition region and coronal lines in the previous studies \citep[e.g.,][]{2011ApJ...738...18T,2011ApJ...736..130T}.
From the above analysis it indicates that the downward and upward propagating features studied in this work are likely plasma flows propagating with opposite directions along the coronal loop legs at the AR boundaries.

The above analysis suggests that we are observing the same process with that from \citet{2012ApJ...749...60M,2021SoPh..296...47T}, i.e. the mass-cycling between the chromosphere and corona.
As indicated in Figure 5 of \citet{2012ApJ...749...60M}, the heated material 
associated with highly dynamic spicules \citep{2009ApJ...706L..80M,2011Sci...331...55D,2014Sci...346A.315T,2019Sci...366..890S} move upward from the chromosphere to the corona with a fast velocity along the coronal loops.
The upflow will lead to a high-speed secondary emission component in coronal lines \citep{2011ApJ...738...18T}. 
The persistent downflows seen in cooler emission are a result of the previously heated plasma transitioning into the radiative cooling domain and then returning to the lower atmosphere with a slower speed.
\textcolor{black}{Also, we noticed that for some events in open loops such as BPD 1, upflows share similar characteristics with self-similar outflows that have been reported by \citet{2023ApJ...955L..38U}. For these events, the mass-cycling between the chromosphere and corona may not be a fully closed cycle and a part of plasma in UPDs can escape outward and into the heliosphere. This process may be a source of fast solar wind \citep{2023ApJ...945...28R,2023ApJ...955L..38U,2022ApJ...933...21K,2023ApJ...943..156K}.}

We noticed that BPDs events can be divided into two categories: downflows only appear in the low-temperature passbands in category\,1, and appear both in the low- and high-temperature passbands in category\,2. This indicates that some downflows only contain cold plasma, and other downflows are mixtures of cold plasma and hot plasma. 
The ubiquitous downflows seen at a temperature of 0.5 MK in transition region observed by SUTRI provide additional evidence for the radiative cooling process in the complex chromosphere-corona mass-cycle. 
In addition, we investigated the relationship between velocities and the radius vector (RV, the fractional distance to the limb from the disk center), to study the center-to-limb variation (CLV) of POS velocities.
RV is defined as, based on \cite{1987ApJ...323..368K,2023ApJ...944..158R}, $\rm RV=\pm\sqrt{x^2+y^2}/R_{sun}$, where $(x,y)$ is the coordinate of the footpoint of the coronal loops in units of arc-second and $R_{sun}$ is the solar radius with a value of 959$^{\prime\prime}$.
\textcolor{black}{A value of zero corresponds to disk center whereas $+1 (-1)$ represents the eastern (western) limb.}
Figure \,\ref{fig:4} plots the POS velocities as a function of RV for downflows and upflows. 
Although the velocities are in wide scatter, we see a weak CLV variation, i.e., larger POS velocities are observed as the RV approaches the limb.
This can be explained by taking into account the LOS projection effect.
Considering the fact that the coronal loops hosting the BPDs are located at the AR boundaries, these loops are likely perpendicular to the line-of-sight (LOS) when the ARs are close to the solar limb.
On the contrary, when the ARs are approaching the solar disk center, the coronal loops are roughly aligned with the LOS and the POS velocities of the BPDs are small.
Such a result and interpretation is consistent with the finding of
\citet{2012ApJ...748..106T}, in which the authors investigated the velocities of the plasma flows from the CLV of the spectroscopic profile asymmetries.

\section{Summary}
\label{sec:sum}
We report 17 BPDs, consisting of coexisting upflows and downflows, from imaging observations taken by SATech-01/SUTRI and SDO/AIA.
These plasma flows persistently appear at the AR boundaries as the ARs rotate from the east limb to the disk center and then to the west limb during the time period of 2022 September 21 to 2022 September 30.

The upflows are prominent in hotter passband, i.e. AIA 193\,\AA, while the downflows are best seen in SUTRI 465\,\AA\ and AIA 131\,\AA.
In some cases, the downflows can also be seen in AIA 304\,\AA, 171\,\AA, and 193\,\AA.
The velocities of downflows are estimated to be only tens of kilometers per second, while the upflows propagate more faster with a velocity of 50-200 \kms.
The coexistence of downflows and upflows can be explained by a chromosphere–corona mass-cycling process, in which the local chromospheric plasma is impulsively heated to the coronal temperature forming an upflow, and then experience a radiative cooling process producing a downflow with the previously heated plasma returning to the lower atmosphere.
\textcolor{black}{For upflows in open loops, some plasma can be ejected into space and this makes upflows a source of solar wind.}
We also investigate the CLV of the POS velocities for both upflows and downflows, which is likely due to the LOS projection effect.

\section{acknowledgement}

This work is supported by the National Key R\&D Program of China (2021YFA0718601), the Youth Innovation Promotion Association CAS (2023061) and the National Natural Science Foundation of China (NSFC, Grant Nos. 12373115). Z. Y. Hou was supported by NSFC grant 12303057 and China Postdoctoral Science Foundation No. 2021M700246.
\textcolor{black}{AIA is an instrument onboard the Solar Dynamics Observatory, a mission for NASA's Living With a Star program.
SUTRI is a collaborative project conducted by the National Astronomical Observatories of CAS, Peking University, Tongji University, Xi'an Institute of Optics and Precision Mechanics of CAS and the Innovation Academy for Microsatellites of CAS.}

\bibliography{sample631}{}
\bibliographystyle{aasjournal}

\end{document}